\documentclass[12pt]{article}
\usepackage{amsmath,amssymb,amsfonts}
\usepackage{graphicx}
\usepackage{xcolor}
\usepackage{float}
\usepackage[T1]{fontenc}
\usepackage{hyperref}
\usepackage{url}
\usepackage[margin=1in]{geometry}

\hypersetup{
  colorlinks=true,
  linkcolor=blue,
  urlcolor=blue,
  citecolor=blue
}

\title{Reconstructive AI Spectroscopy of Charged Particle Beams}

\author{%
  \begin{minipage}{0.9\textwidth}
    \centering
    {Vasily Kozhevnikov}\thanks{Corresponding author: Vasily.Y.Kozhevnikov@ieee.org}, 
    {Andrey Kozyrev}, 
    {Elena Klepalova},\\
    {Victor Tarasenko}, 
    {Evgenii Baksht}\\[4pt]
    \small
    \textit{Institute of High Current Electronics, 634055 Tomsk, Russia}\\
  \end{minipage}
}

\date{\today}

\begin{document}

\maketitle

\begin{abstract}
\noindent This work introduces a physics-informed neural
network (PINN) framework for reconstructing electron energy
spectra from sparsely sampled attenuation-curve data. Leveraging the NVIDIA PhysicsNeMo platform, the proposed mesh-free methodology operates directly on raw experimental datasets
while explicitly incorporating all experimental uncertainties. Validation on subnanosecond electron beam measurements demonstrates that the approach accurately resolves complex, multi-peaked spectral features of energy distribution. The framework enforces physical consistency through embedded governing principles and exhibits substantial predictive capability for energy spectrum reconstruction from noisy, low-precision, and sparse experimental data.
\end{abstract}

\noindent\textbf{Keywords:} electron energy spectrum, PINN, inverse ill-posed problems, NVIDIA \mbox{PhysicsNeMo}, anomalous electrons

\section{Introduction}
The investigation of the energy characteristics of charged and neutral particle beams remains one of the central and most actively pursued problems in accelerator physics \cite{b1}. Of particular interest are the energy spectra of electron beams generated at the output of nanosecond and subnanosecond gas-filled and vacuum diodes when a high-voltage pulse with a short leading edge is applied. Under these conditions, the electron energy spectra typically exhibit complex, multi-peak structures. 

A distinctive feature of subnanosecond discharge systems is the production of electrons with "anomalous" energies $\varepsilon$ that substantially exceed the values determined by the amplitude of the applied voltage $U_0$ across the diode gap ($\varepsilon > qU_0$, where $q$ is the elementary charge). Elucidation of the physical mechanisms responsible for this effect critically depends on the accuracy and reliability of the measured particle energy spectra.

Direct measurement of the spectral energy characteristics of short-duration beams is technically challenging and often yields results of limited reliability \cite{b2}. Consequently, the "attenuation curve method" \cite{b3} has historically served as the principal technique for determining the energy composition of such beams. This method is based on recording the amplitude of the absorbed charge or current as an electron beam traverses metallic filters of different thicknesses. Under contemporary experimental conditions, attenuation curves can be obtained with a relative uncertainty on the order of 1–2~\% or even lower. However, the task of extracting a beam spectrum from a set of $N$ attenuation curve points $g(x),\ x = x_1, x_2 \dots x_N$ is an ill-posed inverse problem for a Fredholm integral equation of the first kind:

\begin{equation}
\int_{\varepsilon_{\min}}^{\varepsilon_{\max}} K(x, \varepsilon) f(\varepsilon) d\varepsilon = g(x)\label{Fredholm}
\end{equation}

\noindent where $f(\varepsilon)$ is the unknown electron energy spectrum to be reconstructed, $K(x, \varepsilon)$ is the kernel of integral equation representing the known attenuation function for a monoenergetic electron beam of energy $\varepsilon$ in a foil of thickness $x$, and $\varepsilon_{\min}$ and $\varepsilon_{\max}$ are the lower and upper boundaries of the energy range considered for the beam. In such problems, variations in the initial data (due to experimental error) lead to unlimited variation in the desired solution. This occurs because the integral operator is compact and its singular values accumulate at zero, so arbitrarily small perturbations in the measured attenuation curve can excite high‑frequency components of the solution with unbounded amplitudes. Consequently, a na\"ive minimization of the residual alone yields an infinite family of mutually distant spectra, all fitting the same experimental data within the error margins. Without additional regularization, these solutions may differ from one another by an arbitrarily large amount, rendering the reconstruction physically meaningless unless prior constraints are imposed.

Over an extended period, the classical Arsenin–Tikhonov algorithm was employed to reconstruct energy spectra of fast electrons in vacuum and gas diodes \cite{b4,b5,b6}. This mathematical method is a variational approach, resulting in a regularization equation that yielded electron energy distributions that are stable with respect to variations in the attenuation curve data \(g(x)\) within the limits of experimental uncertainty. Nevertheless, it exhibited several significant drawbacks that led to computational artifacts in the reconstructed spectrum \(f(\varepsilon)\). The principal limitations of this approach can be summarized as follows (in order of decreasing importance):
\begin{enumerate}
    \item Uncertainties in the kernel of the integral equation, \(K(x,\varepsilon)\), which effectively describe the inaccuracies in modeling the interaction of electrons with the foil material (Tabata–Ito approximation \cite{b7}), were not taken into account;
    \item The fine structure of the reconstructed spectrum was critically sensitive to the preprocessing of the experimental attenuation curve \(g(x)\), in particular to the artificial increase in the number of data points to 50 or more by means of various interpolation schemes;
    \item Tikhonov regularization introduces additional boundary conditions that must be imposed at the edges of the considered energy interval, e.g., $f(\varepsilon_{\min}, \varepsilon_{\max})$, $f^\prime(\varepsilon_{\min}, \varepsilon_{\max})$, $f^{\prime\prime}(\varepsilon_{\min}, \varepsilon_{\max})$, etc. These boundary conditions substantially affect the overall shape of the reconstructed spectrum, although they are only weakly constrained by physical considerations.
\end{enumerate}

In this paper, a fundamentally different computational approach is used. It can be classified as physically-informed neural networks (PINNs). This methodology was introduced first in \cite{b8}. The essence of the approach is that \eqref{Fredholm} is regularized by minimizing a generalized normalized loss function that includes all possible uncertainty propagation, preserving the meshless framework and the ability to operate with raw sparse experimental data. This eliminates the introduction of artificial distortions into the spectrum structure. The purpose of this article is to apply the AI algorithm to reconstruct the spectrum of fast electrons based on a data set of the attenuation curve for a subnanosecond discharge from one of the early experiments \cite{b3}. It is shown that AI analysis makes it possible to verify the energy structure of an electron beam and confirm the existence of its "anomalous" components while fully taking into account experimentally determined uncertainties.

\section{Numerical Method}
As already mentioned above, the full set of uncertainties affecting the structure of the energy spectrum is determined by the experimental error of the right-hand side (attenuation curve) and the kernel of the equation (Tabata-Ito formula \cite{b7})

\begin{equation}
\begin{gathered}
g(x) = \bar{g}(x) \pm \Delta g(x), \\
K(x, \varepsilon) = \overline{\overline{K}}(x, \varepsilon) \pm \Delta K(x, \varepsilon)
\end{gathered}
\label{errors}
\end{equation}

\noindent where $\bar{g}(x)$ and $\overline{\overline{K}}(x, \varepsilon)$ represent the discrete or approximated sets of mean values for the attenuation curve and the integral kernel, respectively. The terms $\Delta g(x)$ and $\Delta K(x, \varepsilon)$ denote the corresponding experimental errors. 

\begin{figure}[H]
\centerline{\includegraphics[scale=0.295]{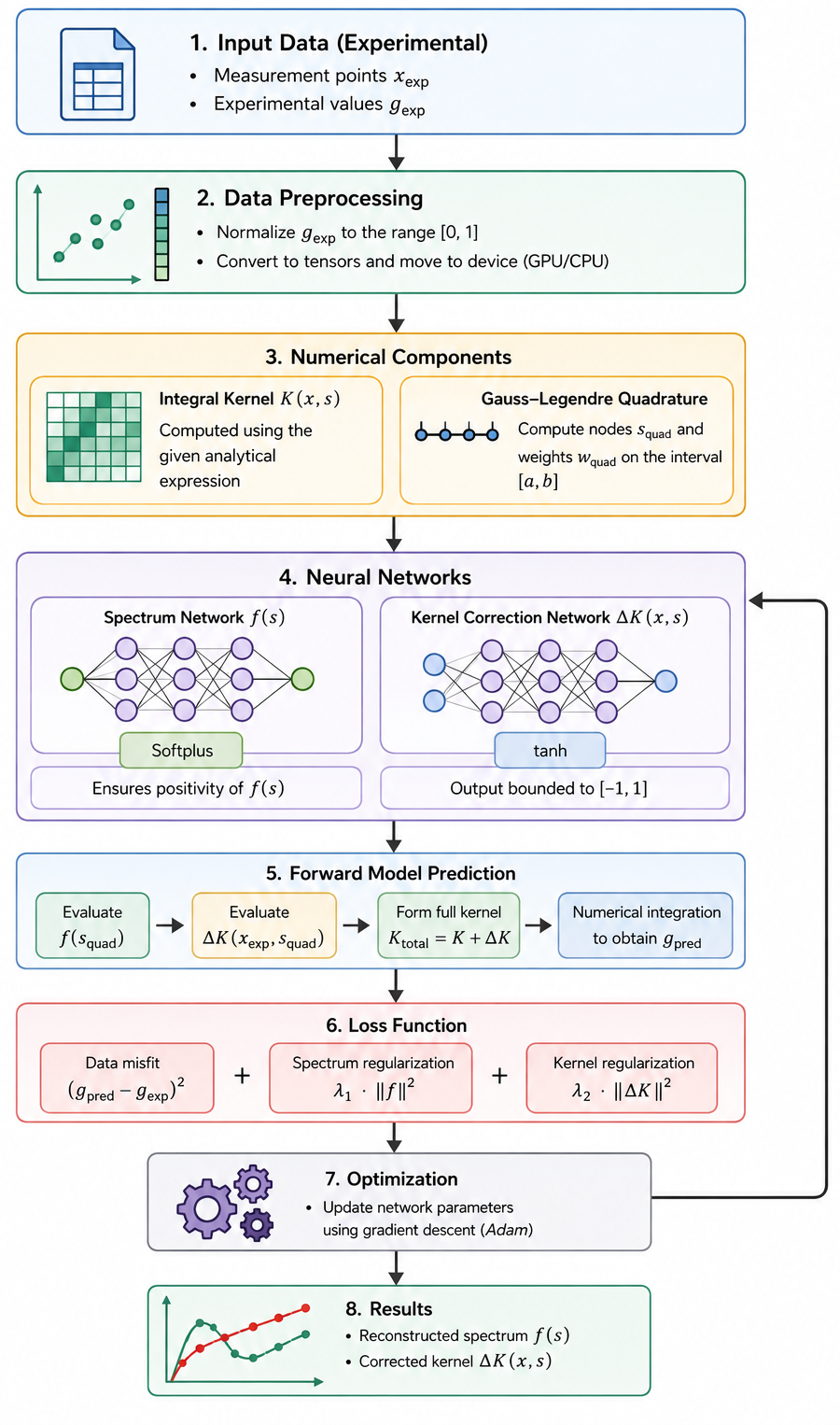}}
\caption{AI-based workflow for solving the inverse Fredholm integral equation \eqref{Fredholm}, including spectrum reconstruction, kernel correction, and regularized optimization.}
\label{fig1}
\end{figure}

For uncertainties \eqref{errors}, we can formulate a generalized normalized loss function. Minimizing it using deep learning methods yields a solution $f(\varepsilon)$ that is robust to all experimental errors in the problem. This normalized loss function $\aleph$ has the form.

\begin{equation}
\begin{split}
\aleph(\theta) = & \frac{\sum_{i=1}^N \left| \int_{\varepsilon_{\min}}^{\varepsilon_{\max}} K(x_i, \varepsilon) f_\theta(\varepsilon) d\varepsilon - g(x_i) \right|^2}{\sum_{i=1}^N |g(x_i)|^2 + \delta} + \\
& + \lambda_1 \frac{\sum_{j=1}^M [f_\theta(\varepsilon_j)^2 + f'_\theta(\varepsilon_j)^2]}{\sum_{j=1}^M |f_\theta(\varepsilon_j)|^2 + \delta} + \\
& + \lambda_2 \left( \frac{1}{MN} \sum_{i=1}^N \sum_{j=1}^M \Delta K(x_i, \varepsilon_j)^2 \right)
\end{split}
\label{loss_function}
\end{equation}

\noindent where $\theta$ denotes the vector of trainable machine learning parameters, including weights and biases, while $N$ represents the number of discrete experimental points and $M$ is the number of nodes in the uniform energy grid used for the regularization term. The parameters $x_i$ and $g(x_i)$ refer to the specific foil thicknesses and their corresponding measured beam charge or current values within the experimental dataset. The constant $\delta$ is introduced in the denominators to ensure numerical stability and prevent division by zero during the normalization of individual loss components. The weighting coefficients $\lambda_1$ and $\lambda_2$ govern the regularization process: the second term in equation \eqref{loss_function} ensures the smoothness of the reconstructed spectrum by penalizing the values of the network-approximated function $f_\theta(\varepsilon_j)$ and its first derivative, computed via automatic differentiation. In contrast, the third term constrains the magnitude of the adaptive kernel correction $\Delta K(x_i, \varepsilon_j)$ generated by the auxiliary neural network. The neural network architecture for the task and the computation process are shown in Fig.~\ref{fig1}. In this methodology, data normalization is not a mandatory procedure.

The computational framework is built upon the NVIDIA~PhysicsNeMo library, utilizing its optimized \textit{FullyConnected} multi-layer perceptron (MLP) components to implement a physics-informed neural network (PINN) architecture. NVIDIA PhysicsNeMo offers a more efficient GPU-accelerated framework compared to the previously used MATLAB Deep Learning Toolbox \cite{b8,b9}. The system consists of two primary interconnected networks: \textit{SpectrumNet}, which approximates the target electron energy spectrum $f(\varepsilon)$, and \textit{DeltaKNet}, which provides an adaptive correction $\Delta K(x, \varepsilon)$ to the integral kernel. \textit{SpectrumNet} is configured with four hidden layers of 128 neurons each using tanh activations, followed by a Softplus output layer to enforce the physical requirement of non-negativity strictly. \textit{DeltaKNet} employs four layers with 256 neurons and a linear output layer (without activation function) to compensate for uncertainties in the theoretical Tabata-Ito model, allowing the correction $\Delta K(x, \varepsilon)$ to take both positive and negative values. 

This transforms the reconstruction into a full inverse problem where the kernel is adaptively adjusted during training. The Fredholm integral operator is evaluated directly within the differentiable framework using a Gauss-Legendre quadrature with 1,000 nodes, ensuring high-order spectral accuracy. The gradients for \textit{SpectrumNet} and \textit{DeltaKNet} are computed simultaneously via backpropagation through the entire computational graph, ensuring consistent adaptation of the spectral reconstruction and the kernel correction without requiring alternating freeze–thaw strategies. Training was conducted using the Adam optimizer with an initial learning rate of $10^{-3}$ and an exponential decay scheduler ($\gamma = 0.99$) over 40,000~epochs. Both networks are trained jointly by minimizing the unified loss function \eqref{loss_function} using a single optimizer (Adam). All computations were performed on an NVIDIA~Quadro~A4000 GPU, leveraging the CUDA toolkit to provide the high-performance throughput necessary for joint network optimization. The overall workflow, including the interconnection between \textit{SpectrumNet}, \textit{DeltaKNet}, and the integral operator evaluation, is schematically illustrated in Fig.~\ref{fig1}.
 
\section{Experiment and Spectrum Reconstruction}
To validate the proposed AI-driven methodology, we utilized experimental data characterizing subnanosecond runaway electron beams generated in a gas-filled diode \cite{b3}. The experiments were conducted in air at atmospheric pressure using the SLEP-150 high-voltage generator, which produces pulses with a voltage amplitude of up to 220~kV and a rise time of approximately 250~ps. The diode featured a sphere-plane geometry consisting of a 9.5-mm-diameter steel spherical cathode and a 10-15~$\mu$m aluminum foil anode, with an interelectrode gap maintained at 8~mm. To refine measurement conditions and limit beam divergence, a copper diaphragm with a 5~mm aperture was frequently employed.

\begin{figure}[H]
\centerline{\includegraphics[scale=0.45]{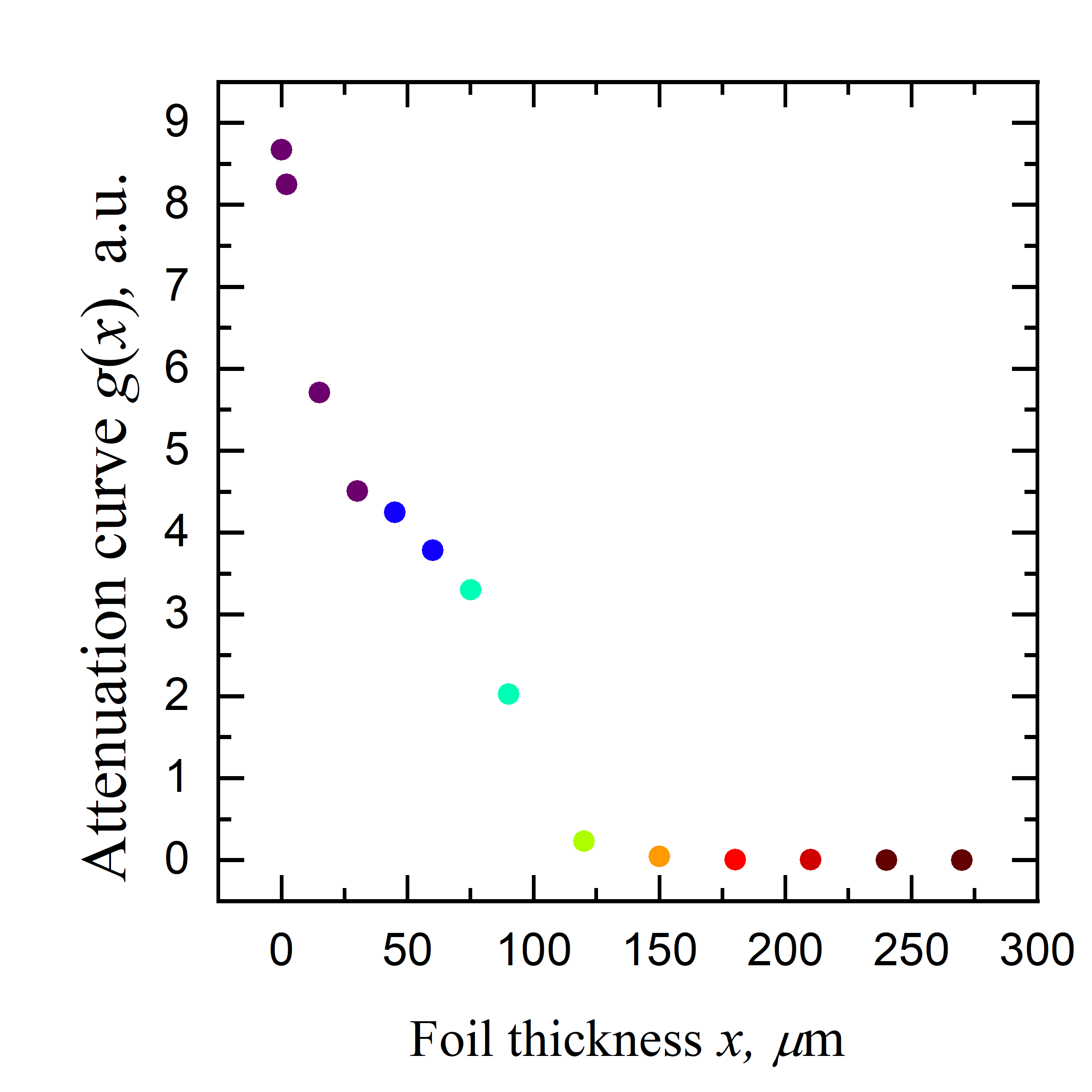}}
\caption{Attenuation curve raw data for a subnanosecond electron beam attenuation in aluminum foils set from experimental paper \cite{b3}.}
\label{fig2}
\end{figure}

The experimental diagnostic suite utilized a high-precision measurement chain specifically engineered to capture the characteristics of subnanosecond electron beams with a temporal resolution of approximately 0.1~ns. The transmitted beam current and total charge were registered by a high-speed collector (Faraday cup) and analyzed using a Tektronix~TDS\mbox{-}6604 digital real-time oscilloscope, which provided a 6~GHz bandwidth and a sampling rate of 20~GS/s to ensure the integrity of ultra-short signals. The primary experimental input (attenuation curve) representing the transmitted beam charge as a function of the thickness of aluminum filters, ranging from 0 to $270~\mu$m, contained 14 raw experimental points \cite{b3}. The energy spectrum was reconstructed in the range from 10~keV to 500~keV using the proposed AI-driven algorithm. For aluminum filters of the given thickness range used in the work, the maximum relative error of the Tabata-Ito formula can be 10-15~\%. At the same time, the accuracy of measuring the attenuation curves (Fig.~\ref{fig2}) does not exceed 5~\%.

\begin{figure}[H]
\centerline{\includegraphics[scale=0.45]{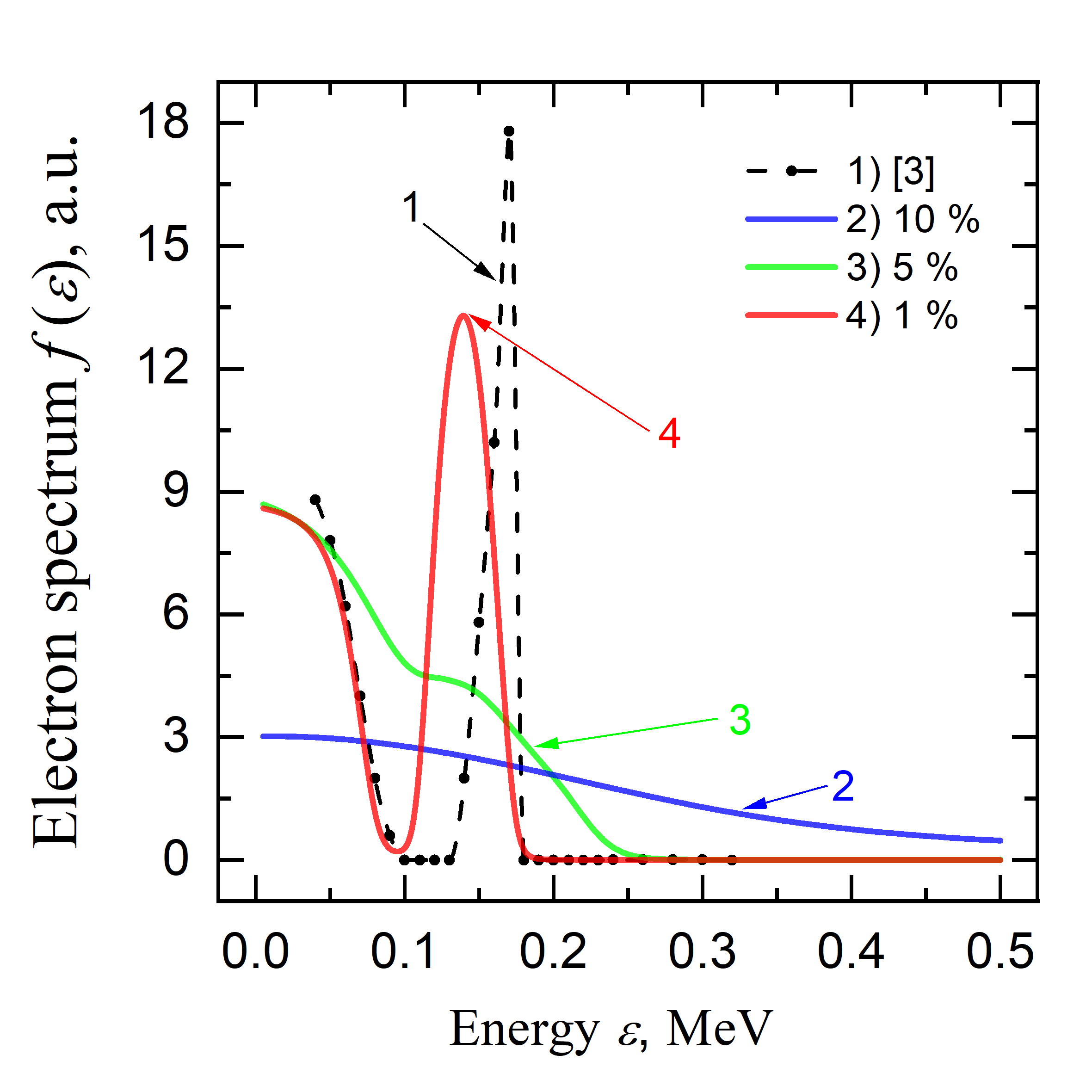}}
\caption{Comparative representation of the energy spectra reconstructed by the AI-based algorithm (curves 2–4) and the experimentally obtained reference spectrum (curve 1) reported in \cite{b3}. The uncertainty of the attenuation curve, expressed as the maximum relative experimental error, is specified in the legend.}
\label{fig3}
\end{figure}

Fig.~\ref{fig3} shows a comparative reconstruction of the electron spectrum behind the anode of the diode of the design under consideration. The dashed line represents the energy spectrum points obtained in \cite{b3}. The colored lines represent the spectrum reconstruction result using the proposed AI-driven reconstruction algorithm. The uncertainty level of the Tabata-Ito formula was not varied, preserving 10~\% for aluminum. At the same time, the reconstruction smoothness and sensitivity to data noise were controlled by the regularization weight $\lambda_1$. Varying $\lambda_1$ effectively changes the regularization strength, allowing exploration of spectral features that are stable with respect to the assumed noise level in the attenuation curve. Lower $\lambda_1$ values reduce the smoothness penalty, enabling the network to resolve finer spectral details, while higher values enforce a smoother solution that is less sensitive to experimental uncertainties. 

It can be seen that reducing the uncertainty value of the attenuation curve leads to the formation of energy spectrum features that are most important for obtaining accurate spectra: the additional maxima and strict non-negativity of the spectrum. Moreover, the AI procedure does not use additional artificially introduced points for the attenuation curve $g(x)$, operating only with the raw experimental data set in Fig.~\ref{fig2}. The method also does not impose any additional conditions on the boundary values of $f(\varepsilon)$, as always required by the Arsenin-Tikhonov procedure \cite{b5,b6}. The main physical observation derived using the current technique is the appearance of a maximum in the region of 130 keV, which corresponds to a group of runaway electrons with anomalously high energies (up to 183~keV). Red curve 4 in Fig.~\ref{fig3} represents the method's predictive horizon, indicating the ultimate spectrum shape for the lowest relative error level $< 1 \%$, which was unachievable experimentally in \cite{b3}. Further increasing the reconstruction accuracy on the data set in Fig.~\ref{fig2} does not lead to any noticeable transformations in the energy spectrum. A more detailed spectrum can be expected to be obtained with a more accurately measured attenuation curve $g(x)$.

\section{Conclusions}
The proposed deep learning–based framework yields a robust and physically consistent solution to the full inverse problem of reconstructing electron-beam energy spectra from experimental attenuation data. By employing a mesh-free formulation that operates directly on raw, sparse experimental datasets, the approach effectively mitigates the numerical artifacts and sampling errors commonly introduced by the interpolation and preprocessing procedures required in classical Tikhonov regularization schemes. Moreover, integrating the \textit{DeltaKNet} architecture enables the model to adaptively compensate for intrinsic uncertainties in the theoretical Tabata–Ito kernel, which may reach up to 15~\%, thereby ensuring a more stable and accurate spectral reconstruction than conventional methodologies.

A principal advantage of this methodology lies in its substantial predictive capability. Even in regimes where the initial experimental accuracy is limited, the PINN architecture can resolve subtle, statistically significant trends embedded within the data. As the presumed fidelity of the input measurements is systematically enhanced via the adjustment of regularization parameters, these latent features—such as the two-peak structure and the population of "anomalous" high-energy electrons—progressively sharpen from broad, qualitative tendencies into quantitatively well-resolved physical structures. 

Beyond its specific application to electron‑beam spectroscopy, the proposed methodology encompasses a broad class of inverse problems that are mathematically equivalent to equation \eqref{Fredholm}, namely Fredholm integral equations of the first kind in which both the right‑hand side and the kernel are affected by experimental uncertainties. Such problems are pervasive in experimental physics and engineering. Representative examples include the reconstruction of radial magnetic‑field profiles in Z‑pinch plasmas from discrete magnetic‑probe measurements, the inference of electron energy distribution functions from bremsstrahlung spectra in high‑temperature plasmas, and the determination of particle‑size distributions from laser diffraction data; all of these can be cast into the same ill‑posed integral framework.

More generally, any diagnostic modality in which the measured quantity is obtained as the solution of a first‑kind Fredholm integral equation \eqref{Fredholm} — whether in plasma physics, accelerator diagnostics, medical imaging, or industrial non‑destructive evaluation—can take advantage of the mesh‑free, uncertainty‑aware framework introduced here. Its capability to operate directly on raw, sparse datasets while explicitly compensating for kernel inaccuracies renders it a versatile and robust methodology, and positions it as a promising basis for a new generation of intelligent data‑processing systems in pulsed‑power science and related fields.

\section*{Acknowledgment}

This paper is dedicated to the cherished memory of Prof.~Dr.~\DJ or\dj e~Paunovi\'c, whose service as Chairman of the TELFOR International Conference Steering Committee has left a lasting legacy in our discipline. We honor him with profound warmth and gratitude—as a mentor who shaped this conference, a senior colleague who inspired us, and a dear friend who shared our journey.

This research has been supported by the program of State assignment of the ISE~SB~RAS, project No. FWRM-2026-0008, FWRM-2026-0009.

\end{document}